\documentclass[referee]{raa}           

\usepackage{graphicx,times}
\usepackage{natbib}
\usepackage{amssymb,amsmath}
\bibpunct{(}{)}{;}{a}{}{,}

\usepackage[pagebackref=true]{hyperref}

\begin{document}

   \title{Map Reconstruction of radio observations with Conditional Invertible Neural Networks}

 \volnopage{ {\bf 20XX} Vol.\ {\bf X} No. {\bf XX}, 000--000}
   \setcounter{page}{1}
   \author{Haolin Zhang
   \inst{1,3}, Shifan Zuo\inst{2}\thanks{E-mail: sfzuo@tsinghua.edu.cn}, Le Zhang\inst{1,3,4}\thanks{E-mail: zhangle7@mail.sysu.edu.cn}
   }
   

   \institute{ School of Physics and Astronomy, Sun Yat-sen University, 2 Daxue Road, Tangjia, Zhuhai, 519082, People's Republic of China; \\
        \and
             Department of Astronomy, Tsinghua University, Beijing 100084, People's Republic of China\\
	\and
CSST Science Center for the Guangdong-Hong Kong-Macau Greater Bay Area, Zhuhai 519082, People's Republic of China\\
\and 
Peng Cheng Laboratory, No.2, Xingke 1st Street, Shenzhen 518000, People's Republic of China\\
\vs \no
   {\small Received 20XX Month Day; accepted 20XX Month Day}
}

\abstract{In radio astronomy, the challenge of reconstructing a sky map from time ordered data (TOD) is known as an inverse problem. Standard map-making techniques and gridding algorithms are commonly employed to address this problem, each offering its own benefits such as producing minimum-variance maps. However, these approaches also carry limitations such as computational inefficiency and numerical instability in map-making and the inability to remove beam effects in grid-based methods. To overcome these challenges, this study proposes a novel solution through the use of the conditional invertible neural network (cINN) for efficient sky map reconstruction. With the aid of forward modeling, where the simulated TODs are generated from a given sky model with a specific observation, the trained neural network can produce accurate reconstructed sky maps. Using the five-hundred-meter aperture spherical radio telescope (FAST) as an example, cINN demonstrates remarkable performance in map reconstruction from simulated TODs, achieving a mean squared error of $2.29\pm 2.14 \times 10^{-4}~\rm K^2$, a structural similarity index of $0.968\pm0.002$, and a peak signal-to-noise ratio of $26.13\pm5.22$ at the $1\sigma$ level. Furthermore, by sampling in the latent space of cINN, the reconstruction errors for each pixel can be accurately quantified.
}

   \authorrunning{Haolin Zhang et al. }            
   \titlerunning{Map Reconstruction of radio observations with Conditional Invertible Neural Networks}  
   \maketitle

\section{Introduction}
Map-making is a critical step in radio astronomy. Before any scientific analysis, it is important to first produce pixelized maps of the observed radio sky from time-ordered data (TOD), with as much accuracy as possible. Mathematically, the reconstruction of sky map from TOD is an ill-posed inverse problem because of observational effects such as scan strategies, noise, complex geometry of the field and data excision due to RFI flagging, etc. There are several map-making methods, with the most common being maximum-likelihood~\citep{Tegmark_1997} that provides the optimal and linear solution. Usually, for solving linear systems in map-making, one use direct methods or iterative methods to achieve the solution. Direct methods are based on brute-force matrix inversion, which requires constructing and inverting the full dense matrix, and is computationally impractical for current computational power when the number of pixels of sky map is more than millions. If the system of equations is singular, then the matrix cannot be inverted, making
the situation even worse. In contrast, iterative optimization methods, such as the commonly-used method of preconditioned conjugate gradients, only require a small memory footprint. However, the number of iterations required to converge to solution could become extremely large and thus the iteration methods can suffer from poor convergence rate. Meanwhile, for ill-posed problems, the derived solution may depend on the choice of the stop criterion of iterations. Additionally, fast gridding methods, such as Cygrid~\citep{Winkel2016} and HCGrid~\citep{Wang2021} with utilizing multiple CPU cores or CPU-GPU hybrid platforms, provide an alternative way for map-making. Although these methods tried to make the most of the hardware, it can not give a map reconstruction uncertainty estimate.

Over recent years, machine learning algorithms, especially those based on deep neural networks, have been widely used in cosmological and astronomical studies and have achieved great success in overcoming many tasks that were previously difficult to accomplish with traditional methods such ~\cite{Lochner:2016hbn,Ravanbakhsh:2017bbi,Schmelzle:2017vwd,Mehta:2018dln,LaPlante:2018pst,Caldeira:2018ojb,Modi:2018cfi,He:2018ggn,Dreissigacker:2019edy,Pfeffer:2019pca,Troster:2019mys,Zhang:2019ryt,Makinen:2020gvh,AIBAORecon...2020arXiv200210218M,2021MNRAS.507.1021N,Wu:2021jsy,2021ApJ...915...71V,2022ApJ...926..151Z,2022ApJ...933..236Z,2022MNRAS.510L...1J,2022arXiv220712511S,2023arXiv230104586W}.

Various neural network methods have been proposed to analyze inverse problems~\cite{2020MNRAS.499.5447K,2022arXiv220200027H,2022EPJC...82..171B,2022MNRAS.512..617K}, and these new data-driven methods demonstrate impressive results. In this study, we will use the multiscale conditional invertible neural network (cINN)~\cite{Ardizzone2019} to solve the ill-posed inverse problem of map-making. Using a FAST-like observation~\citep{Nan2011,Li2013,Li2016,Li2018}, we validate the effectiveness of cINN in map-making and demonstrate such network provides alternative way to reconstruct the sky map from TOD with high-fidelity.

The invertible neural network (INN) was first proposed by~\cite{Ardizzone2018}, and then was soon improved, which is called the conditional invertible neural network (cINN)~\citep{Ardizzone2019}. In order to maintain the unique characteristics of INN, the architecture prohibits the use of some standard components of neural networks, such as batch normalization and pooling layers. Avoiding some fundamental limitations of INN, the cINN combines the INN model with an unconstrained feed-forward network, efficiently preprocessing the conditioning image into the most informative features. Also, cINN allows for the joint optimization of all its parameters using a stable training procedure based on maximum likelihood. This is a new class of neural networks suitable for solving inverse problems. 

cINN originally focus on learning the well-posed forward process (e.g., mapping the true radio sky to TODs), and use additional latent output variables to describe the information lost in the forward process. Due to the invertibility of cINN, the corresponding inverse process is implicitly learned for free through the model. In the specific map-making problem, given a specific observation and the distribution of the latent variables (usually assumed to be Gaussian), the inverse pass of the cINN provides a full posterior distribution over parameter space.  

This study presents a new solution for efficiently reconstructing sky maps by using a conditional invertible neural network (cINN). By generating simulated TODs from a given sky model through forward modeling, which involves drift-scan observations using the FAST configuration, including 19 beams and a frequency range of 1100--1120 MHz. The trained neural network can accurately produce reconstructed sky maps, showing good performance in reconstructing maps from simulated TODs. Moreover, the reconstruction errors for each pixel can be precisely quantified by sampling in the latent space of cINN.

In Sect.~\ref{S:re} we briefly introduce the map-making equations and describe existing methods of map reconstruction and we give a detailed  description of cINN. In Sect.~\ref{S:ex}, we give a description of the simulation and our training data. In Sec.~\ref{S:rs}, we present our results for cINN and demonstrate its good performance in map reconstruction. Finally, we list our conclusions in Sect.~\ref{S:con}.

\section{Methods}\label{S:re}

\subsection{Map-making for single-dish radio telescopes}\label{S:map-making}
Map-making is a crucial step in radio observations, bridging the gap between the collected time-ordered data (TOD) and scientific analysis. For a single-dish radio telescope with a single beam, the map-making input is a series of calibrated TODs, represented by a single time-domain vector $d$ of size $N_t$ containing all antenna measurements. Each measurement at time $t$, $d_t$, is a sum of the sky signal in pixel $p$, $x_p$, and measurement noise, $n_t$, with the beam convolution already applied to the sky signal. The pointing matrix, a sparse and tall ($N_t\times N_p$) matrix, encodes how TOD at each time $t$ responds to each pixel $p$ in the sky map. The TOD is modeled as:
\begin{equation} \label{eq:dt}
  y_{t} = \sum_{p} A_{tp} x_{p} + n_{t},
\end{equation}

or in the matrix-vector form as,
\begin{equation} \label{eq:dv}
y = A x + n,
\end{equation}
where $x$ represents the sky map to be reconstructed. The structure of the pointing matrix $A$ reflects the specific scanning strategy used in the observation.

For observations that involve multiple beams and frequencies, the aforementioned basic model can be expanded as 
\begin{equation} \label{eq:dti}
  y^i_{t}(\nu) = \sum_{p} A^i_{tp} x_p(\nu) + n^i_t(\nu)\,,
\end{equation}
where $\nu$ represents the frequency being observed and the superscript $i$ represents the index of the beam being used.  In the same form as Eq.~\ref{eq:dv}, we can also write the matrix form of the above equation, except that here the matrix and vectors are redefined as $A=[A^1,A^2,\cdots], y=[y^1,y^2,\cdots], n=[n^1,n^2,\cdots]$.

Solving Eq.~\ref{eq:dv} is equivalent to solving a system of linear equations with a large number of parameters, which is a typical linear inverse problem. ~\cite{Tegmark_1997} has proposed a variety of map-making methods, each with its own desired properties. The most common one is the optimal, linear solution, 
$\hat{x}=\big(A^{T} W A\big)^{-1} A^{T} W d$, which is an unbiased estimator for a positive defined weighting matrix $W$. In particular, assuming a Gaussian distributed noise with zero mean and variance of $N$ in the time domain, and choosing the weighting as $W$ = $N^{-1}$, the estimator then becomes the standard generalized least-square solution for the map with minimum variance, 
\begin{equation} \label{eq:mc}
  \hat{x} = H^{-1} b\,,\quad {\rm with}~~H\equiv A^T N^{-1} A\,, {\rm and}~~b \equiv A^T N^{-1} d\,,
\end{equation}
where the noise covariance matrix of the map is $\mathcal{N}=\left(A^T N^{-1} A\right)^{-1}$.  
Since $\big(A^T N^{-1} A\big)$ is generally a dense matrix, a direct brute-force inversion typically costs $\mathcal{O}\big(N_p^3\big)$ flops, which is computationally intractable if $N_p\sim 10^6$ and makes the map-making problem particularly challenging. For noise, since $N$ is sparse in the frequency domain, we need to perform each matrix multiplication on a matrix sparse basis, transforming between the frequency and time domains by using the fast Fourier transform. Furthermore, in practice, the exact matrix inversion may not exist if a matrix is sufficient large of if the matrix is illness or rank-deficient, not sufficient large, leading to solutions that are numerically unstable. And so, one has to use the Moore-Penrose pseudoinverse or some regularization-based methods~\citep{Cheng2011} to approximate the inverse of non-invertible matrices.

More practically, iterative methods have offered a efficient alternative to solve the linear system of map-making, where the class of Krylov methods are involved, such as using preconditioned conjugate gradient algorithm. Explicit inversion of the linear system matrix is avoided by iteratively obtaining successively improved  solutions. The computational complexity of such methods is at most $\mathcal{O}\big(N_p^2\big)$. 

However, a large condition number of the system matrix (the ratio of the largest to the smallest eigenvalue of a matrix) can significantly decrease the convergence rate of iterative solvers, leading to unacceptable time requirements for solutions with required accuracy. Thus one has to carefully choose a preconditioner matrix to the linear system so that the condition number of the preconditioned system becomes much smaller. In practice, the matrix is usually positive semi-defined, which is because the incomplete coverage of pixels in an observed sky area. This incompleteness generally originates from
the choice of scanning strategy and the RFI subtraction in data preprocessing. Therefore, there is a null space such that $Hx=0$, leading to a degeneracy in the estimated sky map $\hat{x}$ (e.g.,~\cite{2010ApJS..187..212C, 2018arXiv180108937P} and references therein). When applying the iterative methods to a semi-defined linear system, the iterative results will start converging towards the optimal solution, and then be hindered to start deviating from the correct solution, because of these degeneracy modes. Therefore, the choice of when to stop iterating is crucial to the successfully solve for such ill-posed map-making problem. Therefore, in order to avoid the aforementioned non-trivial problem, we will propose below a novel deep learning-based approach.

\subsection{Application of neural network to map-making}\label{S:app}

\begin{figure}
  \centering
 \includegraphics[width=0.45\textwidth]{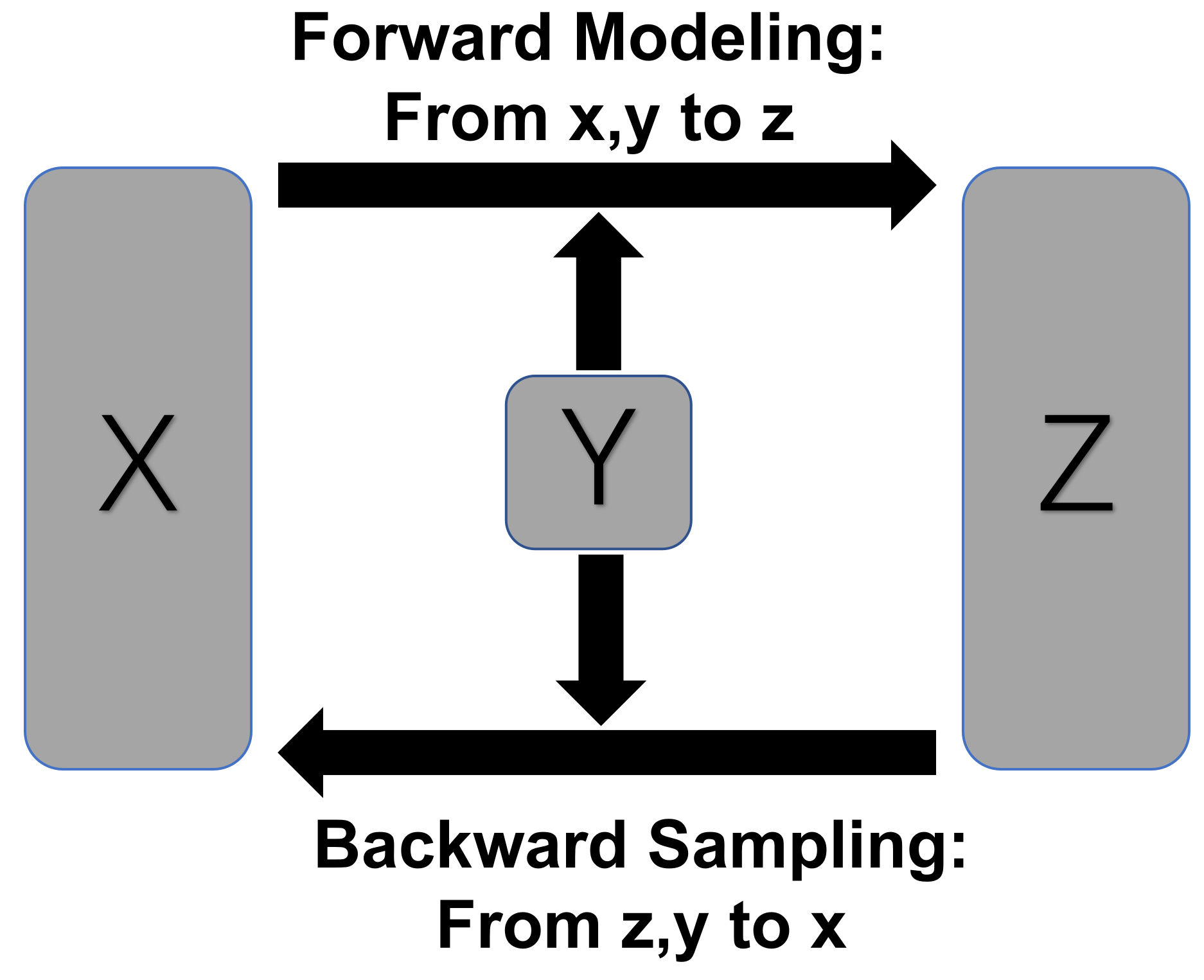}
  \caption{Schematic overview of the conditioned invertible neural network (cINN) for solving the inverse problem of map-making. In the training process, cINN will learn how to transform the pairs $[x,y]$ to latent variables $z$, by optimizing the forward mapping $f(x,y)=z$, where the sky maps $x$ and observational data $y$ (defined as the condition in cINN) are provided by simulations with a known forward modelling as defined in Eq.~\ref{eq:dv}, and serve as inputs to the network. The distribution of the latent variables $p(z)$ is enforced to be Gaussian during the training for simplicity, although $p(z)$ can be arbitrarily assumed.  Due to the invertibility of cINN, the trained network thus provides a solution for the inverse process $f^{-1}$ for free. When making a prediction with a new observation $y$, cINN then transform $p(z)$ to the posterior distribution $p(x|y)$ via the backward mapping $f^{-1}(z,y)=x$. This means the sky map will be reconstructed by sampling the latent variables $z$ drawn from $p(z)$.}
  \label{fig:inn}
\end{figure}

The inverse problem of map-making can be studied under a Bayesian framework. For a given data $y$, the inverse problem of map-making is essentially to derive the posterior distribution, $p(x|y)$, for the true sky map $x$. In the context of mathematics, a forward mapping from any physical parameters $x$ to the associated observed variables $y$, $f(x) \rightarrow y$, is subject to a potential loss of information, which causes degeneracies since $y$ no longer captures all the variance of $x$ entirely. To preserve all information about $x$, the dedicated cINN encodes all variances of $x$ to the latent variables $z$ (unobservable) by learning a mapping from $x$ to $z$, which is conditioned on $y$. Due to the invertible architecture of this network, cINN can provide a solution for the inverse mapping $f^{-1}(z,y) \rightarrow x$ after learning this forward mapping, which is the key point of the cINNs to solve the inverse problem. Thus, such inverse mapping provides an estimate of the posterior distribution $p(x|y)$ by sampling the latent variables $z$. In principle, the distribution of $z$ can be chosen arbitrarily, but for simplicity, we further assume that $z$ follows a Gaussian distribution, enforced during the training process. Fig.~\ref{fig:inn} sketches the concept of the cINN methodology. 

In our case, this means the reconstructed sky map can be automatically retrieved by sampling the Gaussian-distributed $z$ in the latent space via the inverted network ($f^{-1}$),
\begin{equation}
p(x|y)=f^{-1}(z,y)~~{\rm with~~} z \sim p_z(z)=G(z, 0, I_n)\,,
\end{equation}
where $I_n$ is the $n\times n$ unity matrix with choosing $n = dim(z)=dim(x)$.

\subsection{Neural Network Setup}\label{S:nn}
We will describe our new approach for map-making from TODs in this section. Our method employs a neural network architecture based on the conditional invertible neural network (cINN) introduced by~\cite{Ardizzone2018}. The INNs can be constructed easily using the framework for easily invertible architectures (FrEIA) based on pytorch, which is a set of INNs available at~\cite{freia}, without any prior knowledge of normalizing flow. To provide context for the cINN, we will first provide a brief introduction for the invertible neural network (INN), upon which the cINN is based.

\subsubsection{INN architecture} \label{S:inn}
The INNs, discussed in~\citep{Ardizzone2018}, are a type of generative model belonging to the normalizing flow family. This family of models is named normalizing flow because it commonly maps input data from the original distribution to a more standard distribution, usually the normal distribution. Depending on the loss function used, the output distribution can vary. Normalizing flow models encompass a large group of models, but INNs specifically employ affine coupling layers such as RealNVP~\citep{Dinh2016} and GLOW~\citep{2018arXiv180703039K}. Compared with other flow models, INNs have three main advantages: (1) INNs are bijective; (2) the forward and backward mappings in INNs are efficient to compute; and (3) the Jacobian for the forward mapping in INNs is easy to calculate. The architecture of the INN is based on a series of reversible blocks, following the design proposed by~\cite{Dinh2016}.

The input vector, $u$, is divided into two halves, $u_1$ and $u_2$, and these blocks subsequently execute two complementary affine transformations.

\begin{equation}
\begin{aligned}
& v_1=u_1 \odot \exp \left(s_2\left(u_2\right)\right)+t_2\left(u_2\right) \\
& v_{2}=u_2 \odot \exp \left(s_1\left(v_1\right)\right)+t_1\left(v_{1}\right)\,.
\end{aligned}\label{eq:v1u1}
\end{equation}
Here, the use of the element-wise multiplication operator $\odot$ and addition $+$ is employed, where the arbitrarily complex mappings $s_i$ and $t_i$ of $u_2$ and $v_1$, respectively, are represented as any neural networks. These mappings are not mandated to possess inverse functions, as they are evaluated in a solely forward direction.
The inversion of these affine transformations is easily accomplished by following,
\begin{equation}
\begin{aligned}
& u_{2}=\left(v_{2}-t_1\left(v_{1}\right)\right) \odot \exp \left(-s_1\left(v_{1}\right)\right) \\
& u_{1}=\left(v_{1}-t_2\left(u_{2}\right)\right) \odot \exp \left(-s_2\left(u_{2}\right)\right)\,.
\end{aligned}\label{eq:u2v2}
\end{equation}
By introducing the conditional invertible neural network (cINN) as an extension to their original INN method~\citep{Ardizzone2019}, the affine coupling block architecture is modified to include additional conditioning inputs $c$. As the mappings $s_i$ and $t_i$ are only evaluated in the forward direction, even when inverting the network, concatenating the conditioning inputs with the regular inputs of the sub-networks can be done without compromising the invertibility of INNs, e.g., replacing $s_2 (u_2)$ with $s_2 (u_2, c)$ in Eqs.~\ref{eq:v1u1} \& \ref{eq:u2v2}.

\begin{figure*}
  \centering
 \includegraphics[width=0.48\textwidth]{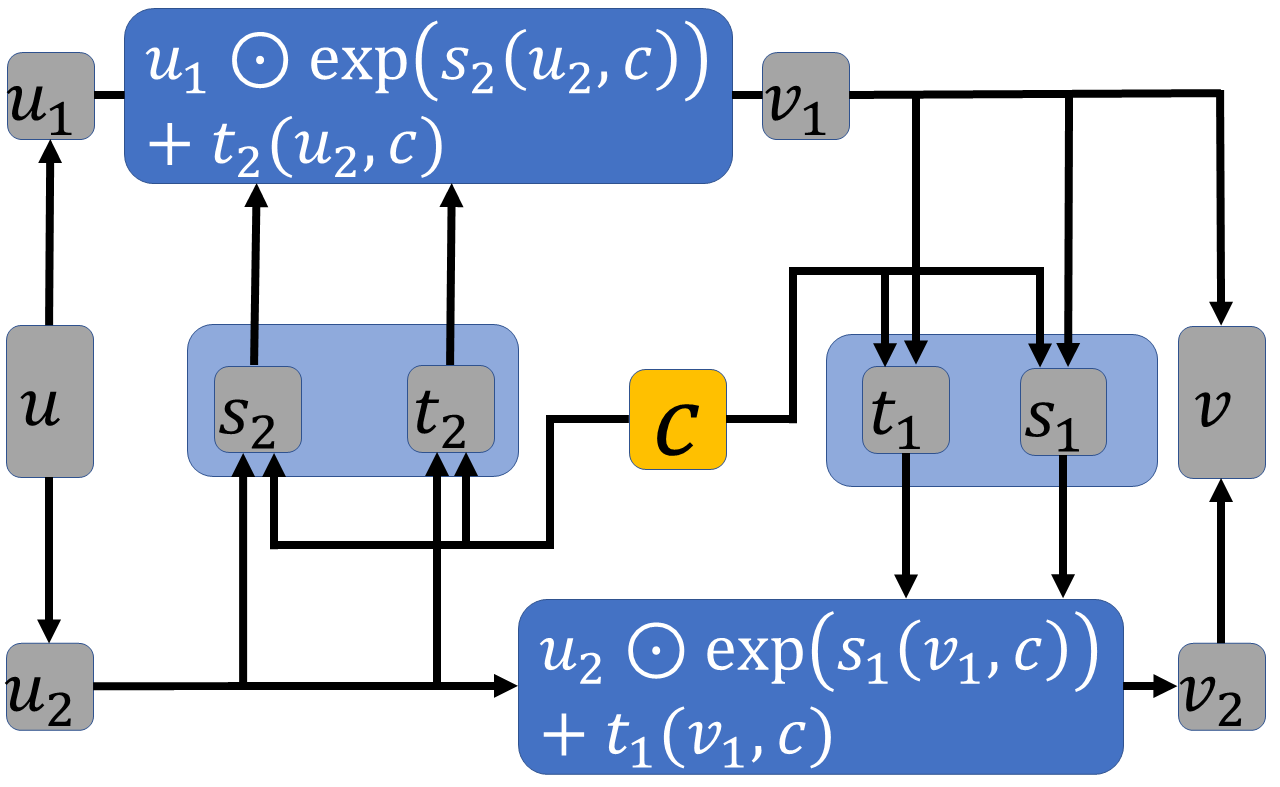}
 \includegraphics[width=0.48\textwidth]{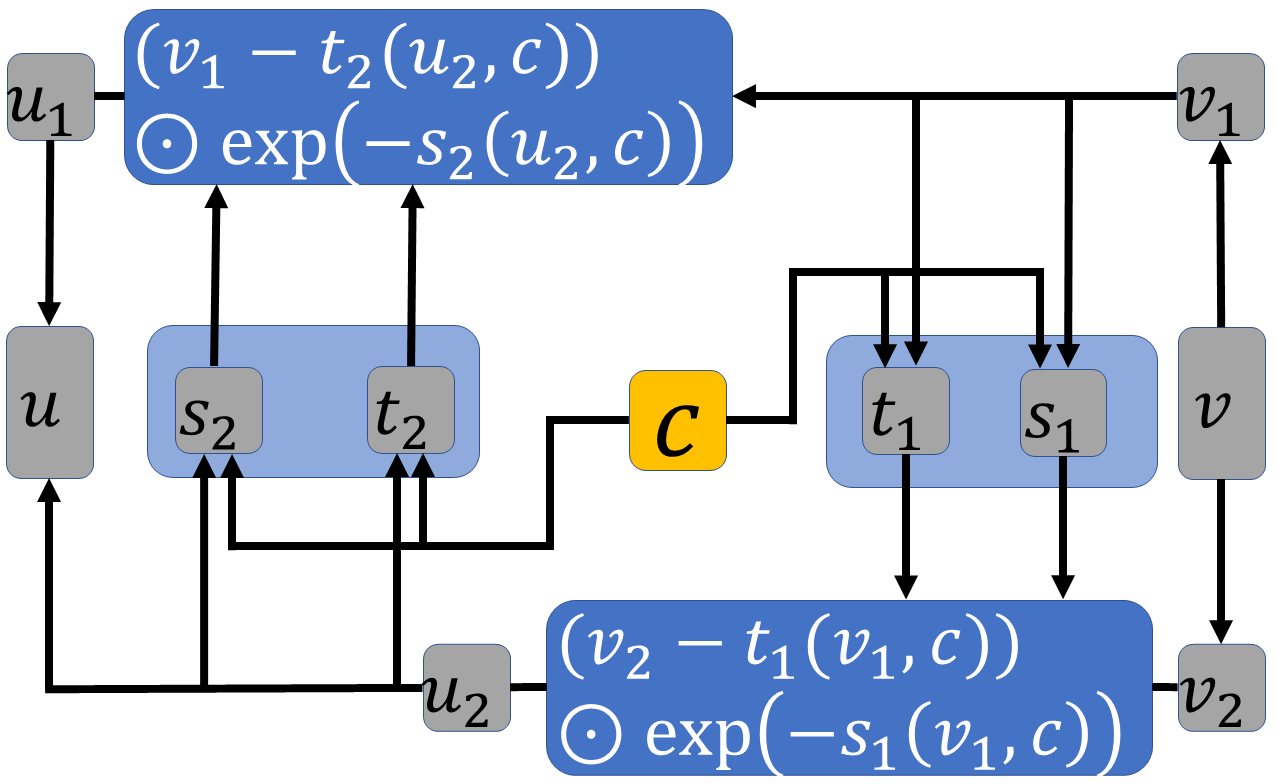}
  \caption{The INN-part of the cINN is formed by stacking multiple coupling blocks, each of which possesses an invertible forward (left) and backward (right) transform through a single conditional affine coupling block (CC). The configuration utilizes a single subnetwork to compute the outputs $s_i()$ and $t_i()$ for each $i$. The left panel illustrates how the data flows through the block in the forward direction (from $x$ to $z$), while the right one displays the inverted case following the affine transformations in Eqs.~\ref{eq:v1u1} \& \ref{eq:u2v2}.}
  \label{fig:innf}
\end{figure*}

\subsubsection{cINN architecture} \label{S:ar}

\begin{figure*}
    \centering
	\includegraphics[width=\textwidth]{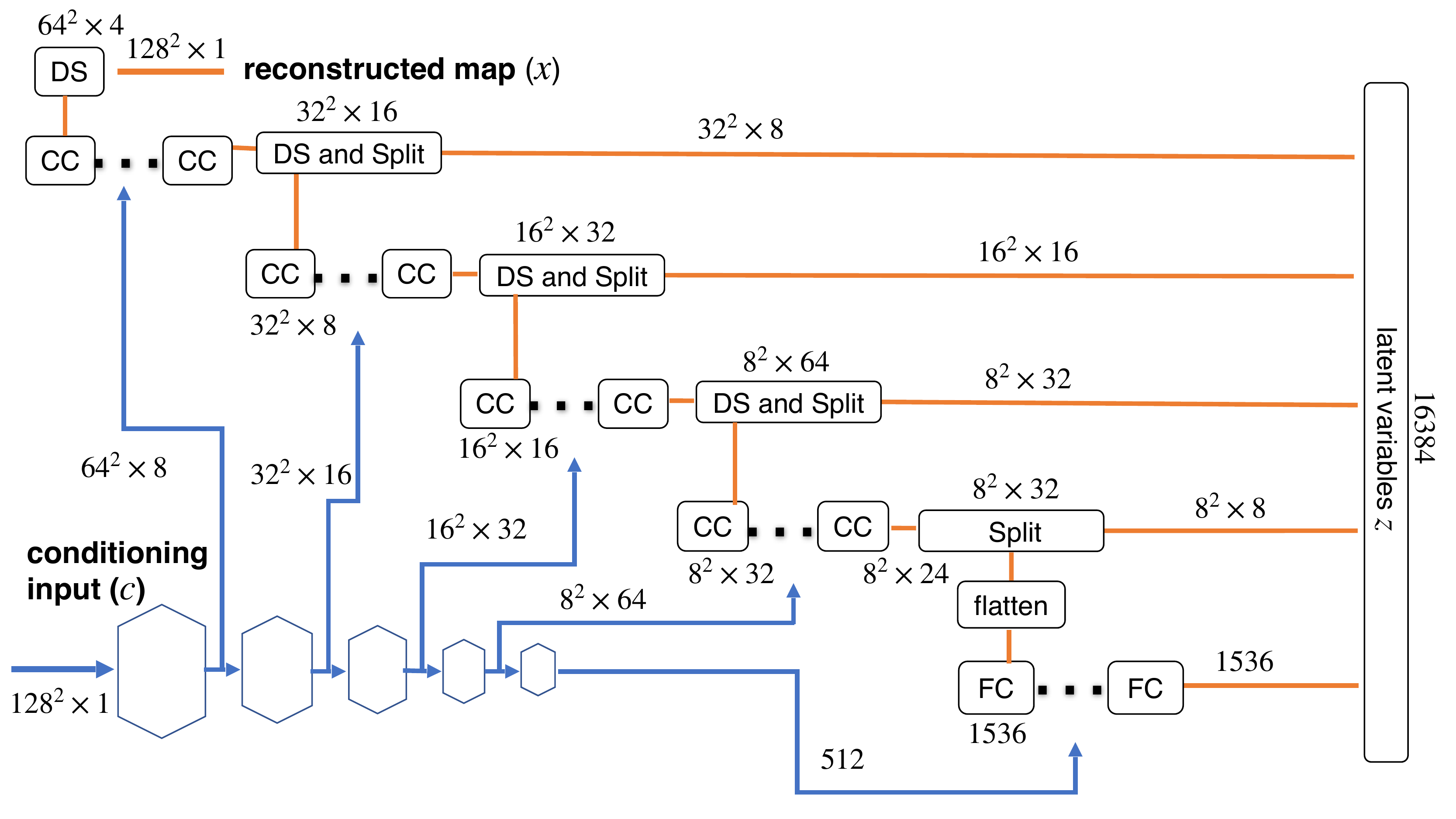}
    \caption{Schematic overview of the entire cINN architecture, employed for the map reconstruction. It resembles the multi-scale cINN presented in~\citet{Ardizzone2019}. There are 16 CC or FC at each level in my network. The conditional coupling block comprises an affine transform layer (shown in Fig.~\ref{fig:innf}) and other invertible layers. The subnet, designated as $s$, $t$, employed in the affine transform layer is a convolutional network suited for mapping data. Meanwhile, the conditional coupling block with fully-connected layer (FC), utilizes an multilayer perceptron neura (MLP) as its subnet, which is suitable for processing vector data. The polygon located in the lower left corner represents the conditional network, which can take the form of any convolutional neural network and serves as the conditional input for the cINN at various levels. The orange and blue lines represent invertible and non-invertible components, respectively.}
    \label{fig:cINN}
\end{figure*}

\begin{figure*}
    \centering
	\includegraphics[width=\textwidth]{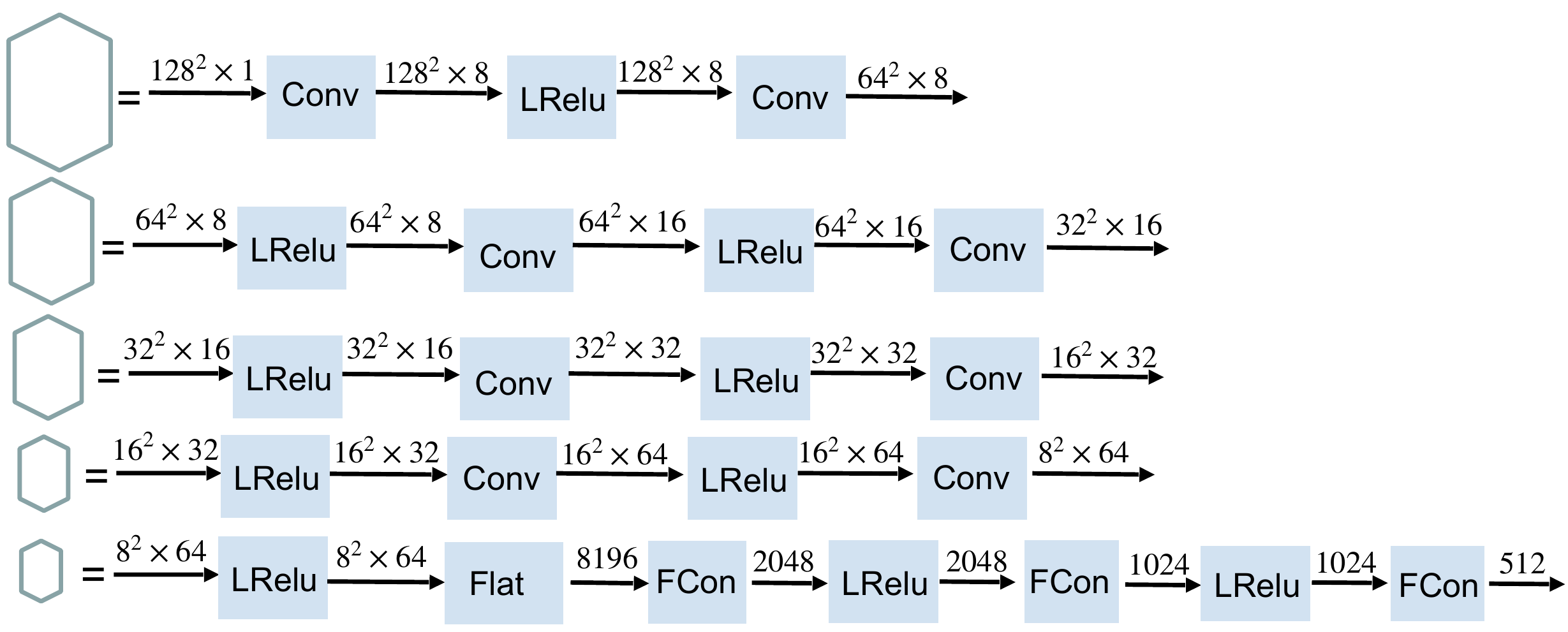} 
    \caption{Details of the conditioning network for the five polygonal components illustrated in Fig.~\ref{fig:cINN}, with using the Convolutional, Fully-Connected, LeakyRelu and Flatten layers. For all convolutional layers, a kernel size of 3 and padding of 1 are used. The stride size is set to 1 for the convolutional layer that preserves the input size, while a stride size of 2 is used for the convolutional layer that changes the size of the input.}
    \label{fig:cINN-1}
\end{figure*}

After the pre-processing stage, the inputs, which are image-like data, are passed through a convolutional network in order to extract features and reduce the computational demands on the INN. The cINN architecture employed for map reconstruction, as depicted in Fig.~\ref{fig:cINN}, is similar to that described by~\citet{Ardizzone2019}. The details of the specific conditioning network (corresponding to the five polygonal components) is provided in Fig.~\ref{fig:cINN-1}. In line with \citet{Ardizzone2019}, the conditional coupling blocks used in this study are from ~\cite{Kingma2018}, which is called GLOW. Each conditional coupling block features a permutation layer that rearranges the channels and facilitates the mixing of information after the affine transformation layer has been applied. Normally, the permutation order remains fixed after initialization. However, GLOW introduces an invertible $1\times1$ convolution layer as a learning permutation layer.

In an ideal scenario, the map resolution remains constant throughout the coupling layers. However, as high resolution maps can be demanding in terms of graphics memory, different resolution stages are employed. Prior to reducing the resolution of the map data, a downsampling layer is employed to decrease its size and increase the number of channels. The downsampling technique utilized is the Haar downsampling, similar to that employed in~\cite{Ardizzone2019}, which is derived from wavelet transforms and is also invertible. The type of downsampling has been found to have a significant impact on training and loss, as noted by~\cite{Ardizzone2019}. A splitting layer is utilized to reduce the dimensionality of the map while increasing its features. Half of the output from each split is concatenated to the latent variable $z$, while the remaining half undergoes further processing through the next coupling block. The choice of the distribution for $z$ can vary, with various distributions being permissible, such as the radial distribution reported in ~\cite{2021arXiv211014520D}. Nevertheless, for the purposes of convenience, the normal distribution is employed as the default choice for $z$ in this study. In the training process, a batch size of 32 is used and the optimizer utilized is Adam with a decay rate of $10^{-5}$.

\subsection{Maximum Likelihood Loss of cINNs} \label{S:mlh}
For the training, a appropriate loss function is required and we refer to ~\cite{Ardizzone2019} for further details on this issue.

The goal is to train a network that establishes a mapping between the distribution in the latent space and the true posterior space of physics parameters. By specifying a probability distribution of $p(z)$, the cINN model $f$ assigns a probability to any input $x$, depending on the network parameters $\theta$ and condition $c$, by means of the probability conservation condition, 
\begin{equation}\label{eq:loss0}
p(x|c, \theta)=p(z)\left|\det\left(\frac{\partial z}{\partial x}\right)\right|\,,
\end{equation}
where $z=f(x|c,\theta)$, and the Jacobian matrix $\partial z/\partial x$ in practice is evaluated at some training sample $x_i$, as $J_i \equiv \det\left(\partial z /\left.\partial x\right|_{x_i}\right)$. Due to the specific structure of the network, the Jacobian is a triangular matrix, which greatly simplifies the calculation of the determinant and ensures its value is non-zero (see details in ~\cite{Ardizzone2019}). Using the Bayes' theorem, $p(\theta|x,c)\propto p(x|c,\theta)p(\theta)$, the optimal network parameters are thus derived by minimizing the loss, which is averaged over $m$ training datasets: 
\begin{equation}
\mathcal{L}=\frac{1}{m}\sum_{i=1}^{m} \left[-\log \left(p(x_i|c_i, \theta)\right)\right]-\log \big(p(\theta)\big)\,.
\end{equation}
Inserting Eq.~\ref{eq:loss0} and adopting the standard normal distribution for variables $z$ for simplicity, i.e., $p(z) =\exp(-z^2/2)$, as well as a flat prior on $\theta$, we obtain the maximum likelihood loss as
\begin{equation}
\mathcal{L}=\frac{1}{m}\sum_{i=1}^{m} \left[\frac{\left\|f\left(x_i| c_i, \theta\right)\right\|_2^2}{2}-\log \left|J_i\right|\right]\,.
\end{equation}

We train the cINN models by minimizing such loss, yielding an estimate of the maximum likelihood network parameters $\theta_*$.  Using this estimate and the inverted network $f^{-1}$, we can then obtain the posterior distribution $p(x|c,\theta_*)$ for a given $c$, by sampling $z$ from the prescribed normal distribution $p(z)$.

\section{Experiments} \label{S:ex}
In this section, the performance of map reconstruction is assessed by utilizing simulated observations. The cINN model is trained until convergence of the maximum likelihood loss is achieved for each training set.

\subsection{Simulated data sets} \label{S:tod}

Here, we present the experimental setup that is implemented to assess the performance of the cINN-based map-making method. The evaluation is conducted using simulated data sets that are modeled after a FAST-like experimental configuration.

\begin{figure}
  \centering

 \includegraphics[width=\textwidth]{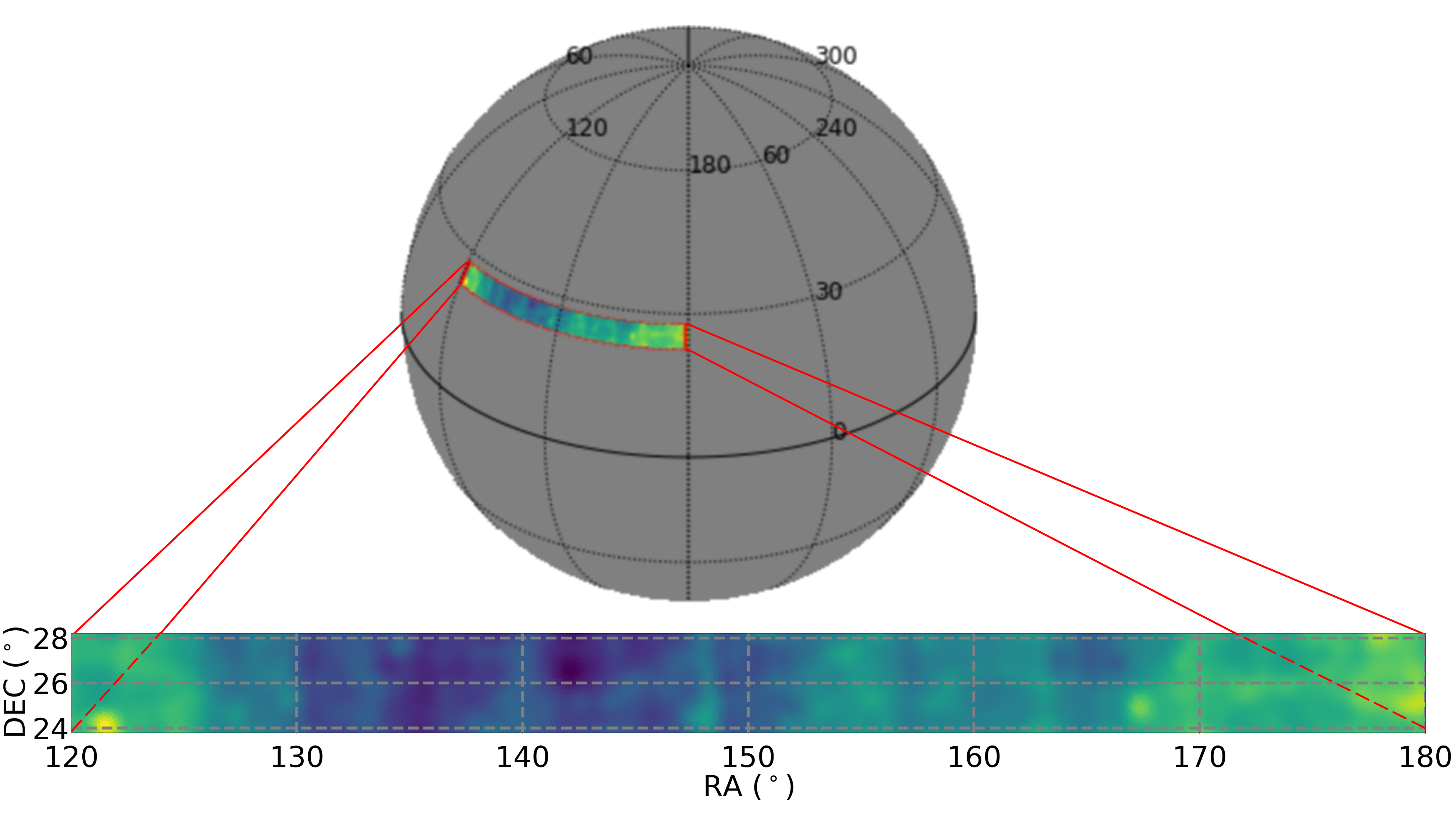}
  \caption{Sky coverage of a drift-scan survey in Equatorial coordinates, using the FAST array consisting of a 19-beam receiver over a 25-day period from May 4th to May 28th. Within the declination (DEC) range of $23^\circ$ to $28^\circ$ and a right ascension (RA) range of $120^\circ$ to $180^\circ$, the survey covers more than 300 square degrees. A zoom-in view of the observed sky is also shown.}
  \label{fig:exr}
\end{figure}

In order to generate TODs by using the forward modelling as described in Eq.~\ref{eq:dv}, we simulated a drift-scan survey using the FAST array consisting of a 19-beam receiver, spanning a period of 25 days, from May 4th to May 28th. The survey covers a sky area of over 300 square degrees within a declination (DEC) range of $23^\circ$ to $28^\circ$ and a right ascension (RA) range of $120^\circ$ to $180^\circ$. The sky coverage is present in Fig.~\ref{fig:exr}. The simulated TODs have a frequency resolution of $\Delta \nu=1$ MHz, in the frequency range of 1100--1120 MHz. With an integration time of 1 s per beam and a total observation time of 14400 s per day, the total number of time samples for all 19 beams amounts to $25\times14400\times20\times19$.

When evaluating the end-to-end performance of an experiment, it is important to simulate correlated noise components, but in this study we focus only on the performance of the cINN, which depends on the mapping matrix $A$ constructed by the scanning strategy, the beam response and noise. We thus assume that the TODs are well-calibrated, meaning that the low-frequency $1/f$ noise in the time streams has been completely filtered out and any other non-ideal instrumental effects such as RFIs and standing waves are not considered in our simulations. As a result, the noise in the TODs is comprised of only white noise. The white noise level in the time streams is proportional to the system temperature $T_{\rm sys}$, and the standard deviation of the noise can be calculated as follows for a given band width $\Delta \nu$ and integration time $\tau$, 
 
\begin{equation}
\sigma_{N}=\frac{T_{\rm sys}}{\sqrt{\Delta \nu \tau}}\,.
\end{equation}

To train our cINN model, we generate various time-ordered data (TOD) at different noise levels by altering the value of $T_{\rm sys}$ randomly from 0 to 25 K in 1200 realizations, with reference to the Fast-like survey. By using the HEALPix pixelization scheme with $N_{\rm side}=512$ for the simulation, the resulting noise levels typically yield noise standard deviations ranging from 0 to 9 mK per pixel, with an angular resolution of 6.87 arcmin.  Based on the FAST configuration, the TOD simulations are performed using Equatorial coordinates for the maps convolved with a Gaussian beam, where the full width at half maximum (FWHM) is slightly frequency dependent, ranging from 4.506 to 4.584 arcmin in the frequency interval of interest.

 Moreover, the simulated true sky map $x$ consists of several Galactic diffuse components such as the synchrotron and free-free emissions, and bright point sources, which are produced from the GSM model~\citep{Costa2010,Zheng2017} and the NVSS catelog~\citep{Condon1998}, respectively.

Additionally, to produce sufficient data samples for training the cINN model, we also employ data augmentation in pre-processing through straightforward techniques, such as randomly rotating sky patches and utilizing different noise realizations. Upon convergence of the maximum likelihood loss during training, we assess the performance of the trained cINN model on the training data.

\subsection{Data pre-possessing} \label{S:tr}
In preparation for training the cINN for map reconstruction. The input to the network is a 2D map of the observed sky region ($x$), which is interpolated from sky map using the appropriate pixelization scheme, HEALPix. The condition ($c$) for the cINN is usually represented by the observed TODs ($y$), but due to the ToDs' varying length and large data size, preprocessing is needed before they can be fed into the network. A more convenient alternative is to use a related quantity with the same length as the sky map. For testing purposes, a gridding map (let $c=y_{\rm grid}$) is used, obtained by assigning TODs to their closest grid points using the \texttt{histogram2D} function in \texttt{numpy}, which is considered as a coarse, but simple and efficient gridding method. 
For simplicity, the TODs are gridded onto 2D flat-sky maps, with each map having an area of $4.3\times4.3$ square degrees and a resolution of $128\times128$. Subsequently, the resulting reconstructed maps also possess the same resolution.

We randomly select 240 observations from different position and system temperature as our data set, each consisting of 20 frequency channels and 5 different realizations. Thus, there are totally $240\times20\times5=24000$ pairs of samples, each sample consisting of a pair of the true sky map and a resulting gridding map from TODs, specifically, represented as $\big(\big[x^i=x^i_{\rm true}, c^i=y^i_{\rm grid}\big]\big)$ for the $i$-th sample. For the purpose of training the cINN, 20,000 samples are utilized, with 2,000 samples being reserved for validation and an additional 2,000 for testing. The training of the cINN model is performed on a GPU server.

\section{Results and Discussion} 
\label{S:rs}
\subsection{Evaluation metrics}
In order to determine the performance of the cINN model in map reconstruction, it is necessary to compare the reconstructed map $x_{\rm rec}$ with the actual map $x_{\rm true}$ using suitable metrics. In the following, we shall introduce several such metrics. One such metric that is commonly used is the mean square error (MSE), as defined by
\begin{equation} \label{eq:MSE}
  {\rm MSE}\big(x_{\rm true}, x_{\rm rec}\big) = \frac{1}{N}\sum_{k=1}^{N}\big(x^k_{\rm true}-x^k_{\rm rec}\big)^2\,, 
\end{equation}
which is calculated by averaging over all pixels of the maps and provides a direct measurement for the mean of the squares of reconstruction error.

Moreover, the Peak Signal-to-Noise Ratio (PSNR) is adopted as a means of evaluating the reconstruction quality~\cite{5596999}. This metric, which is expressed as a log-scaled MSE, can be represented as follows:
\begin{equation} \label{eq:PSNR}
  {\rm PSNR}(x_{\rm true}, x_{\rm rec}) = 10 \log_{10}\left( \frac{L^2}{{\rm MSE}\big(x_{\rm true}, x_{\rm rec}\big)}\right)\,.
\end{equation}

where $L$ is a scalar chosen to reflect the dynamic range of the ground truth map. In this study, $L$ is defined as the difference between the maximum and minimum values in the true map, $L=|x_{\rm max} - x_{\rm min}|$.  Essentially, a higher PSNR value is indicative of improved reconstruction accuracy.

Additionally, the structural similarity index measure (SSIM)~\citep{1284395} is used to evaluate the overall structural similarity between the true and reconstructed maps. It aligns with human visual perception of similarity and the values are in the range of $[0, 1]$, with higher values indicating better performance. The value of SSIM is calculated through
\begin{equation} \label{eq:SSIM}
  {\rm SSIM}(x_{\rm true}, x_{\rm rec}) = \frac{(2\mu_i\mu_j + C_1)(2\Sigma_{ij} + C_2)}
  {(\mu^2_{i} + \mu^2_{j} + C_1)
  (\sigma^2_{i} + \sigma^2_{j} + C_2)}\,,
\end{equation}
where $\mu$ and $\sigma$ represent the mean and variance of a map, respectively, with $i$ and $j$ denoting the true and reconstructed map. $\Sigma_{ij}$ refers to the covariance between the two maps. The positive constants $C_1=(k_1 L)^2$ and $C_2=(k_2 L)^2$ are included to prevent a null denominator and to stabilize the calculations, with values of $k_1=0.01$, $k_2=0.03$.

\subsection{Results of map reconstruction} \label{S:mrr}

The main advantage of the cINN framework lies in its ability to efficiently estimate the full posterior of the reconstruction on a pixel-by-pixel basis, enabling effectively capture the underlying probabilistic relationships between the observed data and the reconstructed map. The large number of reconstructed maps helps to account for the inherent uncertainty and variability in the data, yielding a more robust and accurate representation of the posterior distribution. 

Based on our tests, we have found that generating 200 maps via sampling $z$ only takes less than 1 second using a typical graphics card. This is a remarkable speed, considering that it involves the estimation of a total of $16,384$ posteriors for each map, given the map resolution of $128\times128$.

\begin{figure}
    \centering
	\includegraphics[width=0.6\textwidth]{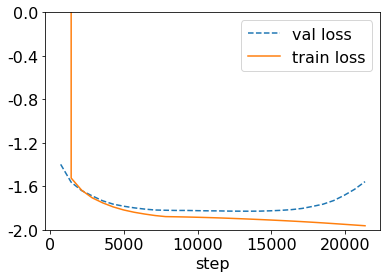}
    \caption{Loss function of map reconstruction as a function of training steps. We stop training at step 15000, where the validation loss reaches its minimum.}
    \label{fig:loss_graph}
\end{figure}
In Fig.~\ref{fig:loss_graph}, the changes in the loss function over training steps are presented. It is well-known that using a low learning rate results in a gradual decrease in loss due to slow parameter updates, while a high learning rate may hinder the search for a solution. Conversely, a high learning rate can prevent finding a solution. Consequently, a learning rate of 0.001 is selected for map reconstruction in this study. 

We have used the initialization technique mentioned in \cite{Ardizzone2019}, where Xavier initialization~\citep{pmlr-v9-glorot10a} is used and the parameter values in the last convolutional layer of sub-networks $s$ and $t$ are set to zero. However, there is still a certain probability that the training is unstable. In our experiment, we encountered a situation where the loss quickly diverged at the beginning. We selected different random seeds and recorded the random seed that would not diverge at the beginning as the initial value. In this way, the loss will have the same downward trend as shown in Fig~\ref{fig:loss_graph}. As observed, the training loss (orange curve) and validation loss (blue curve) are both minimized when using this learning rate. However, after step 15,000, the validation loss began to increase, even though the training loss continued to decrease. Continuing to train the cINN model, even if the training loss continues to decrease, may result in a significant rise in validation loss. Therefore, we stop training at this step, where the validation loss reaches its minimum.

To further investigate the effects of underfitting and overfitting on the map reconstruction, we have chosen two checkpoints above and two below the currently selected one, say steps 4,000 and 20,000. Our findings show that for the underfitting case, MSE  is about $(3.9\pm19.3)\times 10^{-4}~\rm K^2$, SSIM is $0.92\pm0.004$, and PSNR is $22.5\pm2.4$ for the test samples. For the overfitting case, MSE is $(4.2\pm4.4)\times 10^{-4}~\rm K^2$, SSIM is $0.96\pm0.004$, and PSNR is $24.4\pm6.2$. Both overfitting and underfitting result in significantly high MSE values when compared with the MSE values listed in Tab.~\ref{table:t1}. In particular, the underfitting also dramatically increases the statistical uncertainty in the MSE values. Furthermore, neither case results in a substantial improvement in the SSIM and PSNR values. Consequently, the MSE metric seems to be more sensitive to the quality of the reconstruction, and the optimal results are achieved by the currently selected point. Additionally, we have to mention that the checkpoints located near the bottom of the loss require careful selection and evaluation, based on the trends observed in the training and validation loss.

\begin{figure*}
    \centering
	\includegraphics[width=0.9\textwidth]{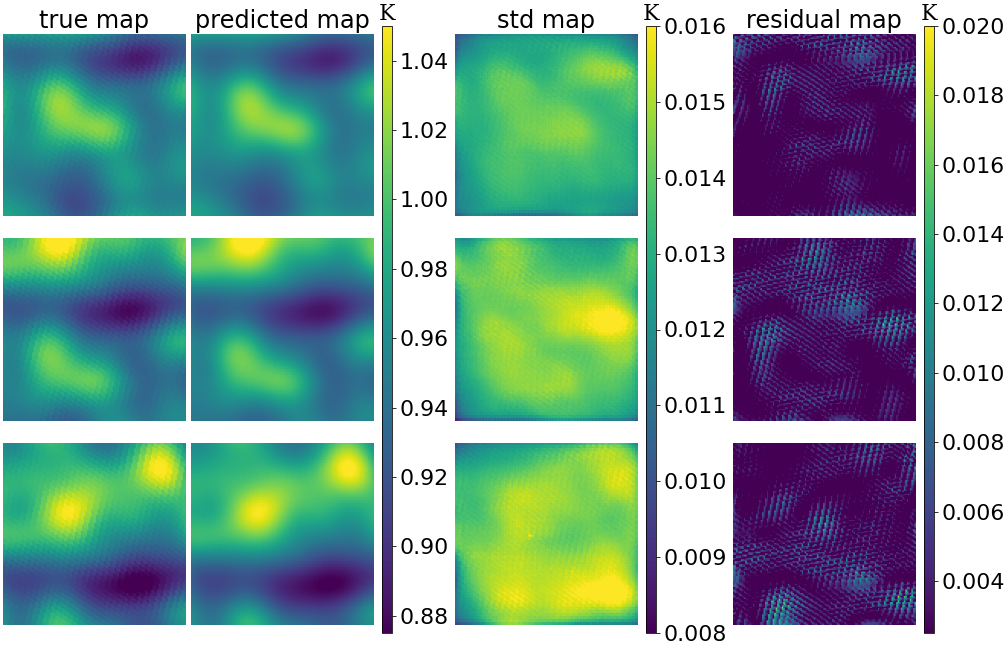}
    \caption{Comparison of the predicted map from the trained cINN model with the original ground-truth map, in units of K, where 3 examples of randomly selected observation samples $y_{\rm grid}$ (from top to bottom) are used as conditioning $c$ inputs for cINN. To demonstrate the reconstruction quality, from left to right, we show the true and predicted maps, the standard-deviation map which represents the uncertainty of the predicted map, and the residual map which shows the deviation between the true map and the mean of predicted maps.}
    \label{fig:reconstruct_map}
\end{figure*}

As shown in Fig.~\ref{fig:reconstruct_map}, each row of panels from left to right represents the original, predicted, standard-deviation, and residual maps, respectively. Here, for a given observation ($y_{\rm grid}$), the trained cINN model obtains the posterior distribution $p(x|y)$ for each pixel by generating 200 reconstructed maps through drawing latent variables $z$ from a prescribed normal distribution. Thus, the predicted map and the standard-deviation map are estimated from the mean and associated $1\sigma$ error of posteriors, indicating the average and the uncertainty in the reconstruction. Moreover, the residual maps demonstrate the deviation level between the mean estimate and the truth in the reconstruction process. As seen, the cINN reconstruction appears to be of good quality, as evidenced by the standard deviation and residuals, which are typically around 0.01 K, the same level as about 1\% of the true map.

\begin{figure*}
    \includegraphics[width=0.32\columnwidth]{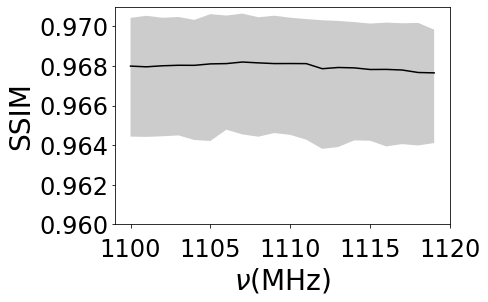}
     \includegraphics[width=0.32\columnwidth]{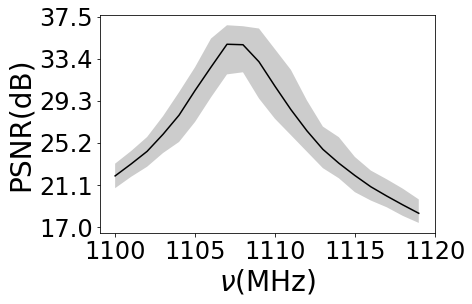}
     \includegraphics[width=0.30\columnwidth]{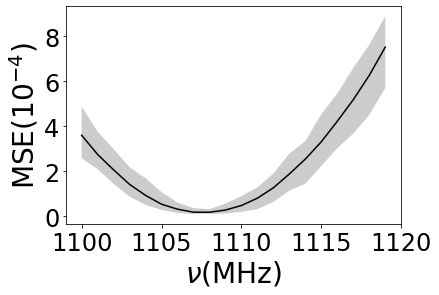}
    \caption{Reconstruction accuracy for different frequency bins measured using three different ways. The mean (solid black) and associated $2\sigma$ uncertainty (gray shaded) are estimated from the entire set of test samples. SSIM (left) measures the overall structural difference between the two maps, better capturing the human perceptual understanding of the difference between two maps. PSNR (middle) and MSE (right) evaluate the reconstruction quality using the signal-to-noise ratio and the absolute difference in image pixels, respectively. Note that for SSIM and PSNR, larger values correspond to a better image reconstruction, while for MSE, the opposite is true.}
    \label{fig:MSE}
\end{figure*}

The values of the three metrics, MSR, SSIM and PSNR, for 20 frequency bins are presented in Fig.~\ref{fig:MSE}, where the mean and 2$\sigma$ uncertainty are estimated from the entire set of test samples. We observe the values of SSIM consistently close to 0.968 across all frequencies, with little variations of approximately 0.001 ($2\sigma$ uncertainty). This suggests that, 1) the structural similarity remains relatively stable and does not significantly change as the frequency varies; 2) the reconstructed maps closely resemble the true ones in terms of structural similarity, and the quality of the reconstructed maps is high. Furthermore, the PSNR values indicate that the reconstructed maps exhibit a relatively low MSE, with a range of 17 to 35 dB. In comparison with the typical temperature of 1 K for true maps, the MSE values range from approximately $1\times 10^{-4}$ to $7\times 10^{-4}$ $\rm K^2$ across all frequencies, which also demonstrates a high quality in map reconstruction.

The results of these metrics averaged over all frequencies and test samples, reported in Tab.~\ref{table:t1}. Specifically, the mean MSE value of $2.29\times 10^{-4}~{\rm K}^2$ indicates that the reconstructed maps have a low level of deviation from the truth in terms of pixel-level accuracy. The mean SSIM value of 0.968 again indicates a high level of structural similarity. Finally, the mean PSNR value of 26.13 dB indicates that the reconstructed maps do not deviate significantly from the true maps in terms of image details. Overall, these metrics suggest that the reconstructed maps highly agree with the true maps.

\begin{table}
\begin{center}
\caption{Map reconstruction performance for the cINN models we have trained. The average values of MSE, SSIM and PSNR and associated 1$\sigma$ statistical errors are shown, respectively, calculated across all frequencies and the entire test samples.}
\label{table:t1}
\begin{tabular}{cccc}
\hline\hline
Performance   & MSE ($\times 10^{-4}$ $\rm K^2$)  & SSIM   & PSNR \\
       \hline
      & $2.29 \pm 2.14$ & $0.968 \pm 0.002$ & $26.13 \pm 5.22$\\
\hline\hline
\end{tabular}
\end{center}
\end{table}

\subsection{Posterior distributions for the map reconstruction} \label{S:domr}

In order to determine the accuracy of the predicted posterior distributions for the map reconstruction at each pixel, we calculate the median calibration error $e^{\rm med}_{\rm cal}$ given in ~\citep{2020MNRAS.499.5447K} and ~\citep{Ardizzone2019}. The correct shape of the posterior distribution is reflected by the calibration error, which makes it a significant evaluation metric for the network. For a given confidence interval $q$, the calibration error is computed over a set of $N$ observations as
\begin{equation} \label{eq:ecal}
  e_{\rm cal} = q_{\rm in}-q\,.
\end{equation}
Here $q_{\rm in}=N_{\rm in}/N$, the fraction of observations that fall within the $q$-confidence interval of the corresponding predicted posterior distribution by cINN. A negative value of $e_{\rm cal}$ indicates that the model is overconfident, meaning that it predicts posterior distributions that are too narrow. Conversely, a positive value of $e_{\rm cal}$ suggests that the model is under-confident, implying that predicts posterior distributions that are too broad. We compute $e^{\rm med}_{\rm cal}$ as the median of the absolute values of $e_{\rm cal}$ across the confidence range of 0 to 1, in steps of 0.01. In addition, the other quantity for evaluation is a median uncertainty interval at a 68\% confidence level, $u_{68}$, corresponding to the $\pm1\sigma$ width of the posterior distribution for the given confidence interval, where we determine the median value over the entire test set.

Using the metrics of calibration error and the median uncertainty interval at 68\% confidence, the results are presented in Tab.~\ref{table:t2}. One can find that the median calibration error falls within the range of about 0--6\%, indicating that the model has relatively high accuracy. In terms of $u_{68}$, our cINN model yields a typical error value of 0.03. Thus, This value is comparable to $\sqrt{\rm MSE}$ ($\sim 0.01$ K), implying a relatively considerable degree of uncertainty in the parameter being estimated. Despite this broadness, the typical error value is still considered remarkably low and acceptable. Given the large number of parameters involved, namely $128\times 128$ for each frequency map, the achievement of this level of performance is particularly remarkable. Thus, we conclude that the performance of our cINN model is sufficient and meets the requirements for our intend application.

\begin{table}\centering
\setlength\tabcolsep{14pt} 
\caption{Performance of our trained cINN model on reconstruction at 9 randomly selected pixels. The results are presented in terms of the calibration error $e_{\rm cal}$ and median uncertainty at 68\% confidence level $u_{68}$ (i.e. the width of a 68\% confidence interval).}
\begin{tabular}{ccc}
\hline \hline pixel index  & $e_{\rm cal}^{\rm med}$ & $u_{68}$ \\
\hline$(32,32)$ & 0.057 & 0.029 \\
\hline$(64,32)$ & 0.041 & 0.030 \\
\hline$(96,32)$ & 0.016 & 0.032 \\
\hline$(32,64)$ & 0.046 & 0.029 \\
\hline$(64,64)$ & 0.043 & 0.030 \\
\hline$(96,64)$ & 0.012 & 0.033 \\
\hline$(32,96)$ & 0.054 & 0.029 \\
\hline$(64,96)$ & 0.025 & 0.031 \\
\hline$(96,96)$ & 0.001 & 0.033 \\
\hline\hline
\end{tabular}\label{table:t2}
\end{table}

\begin{figure*}
    \centering
	\includegraphics[width=0.75\textwidth]{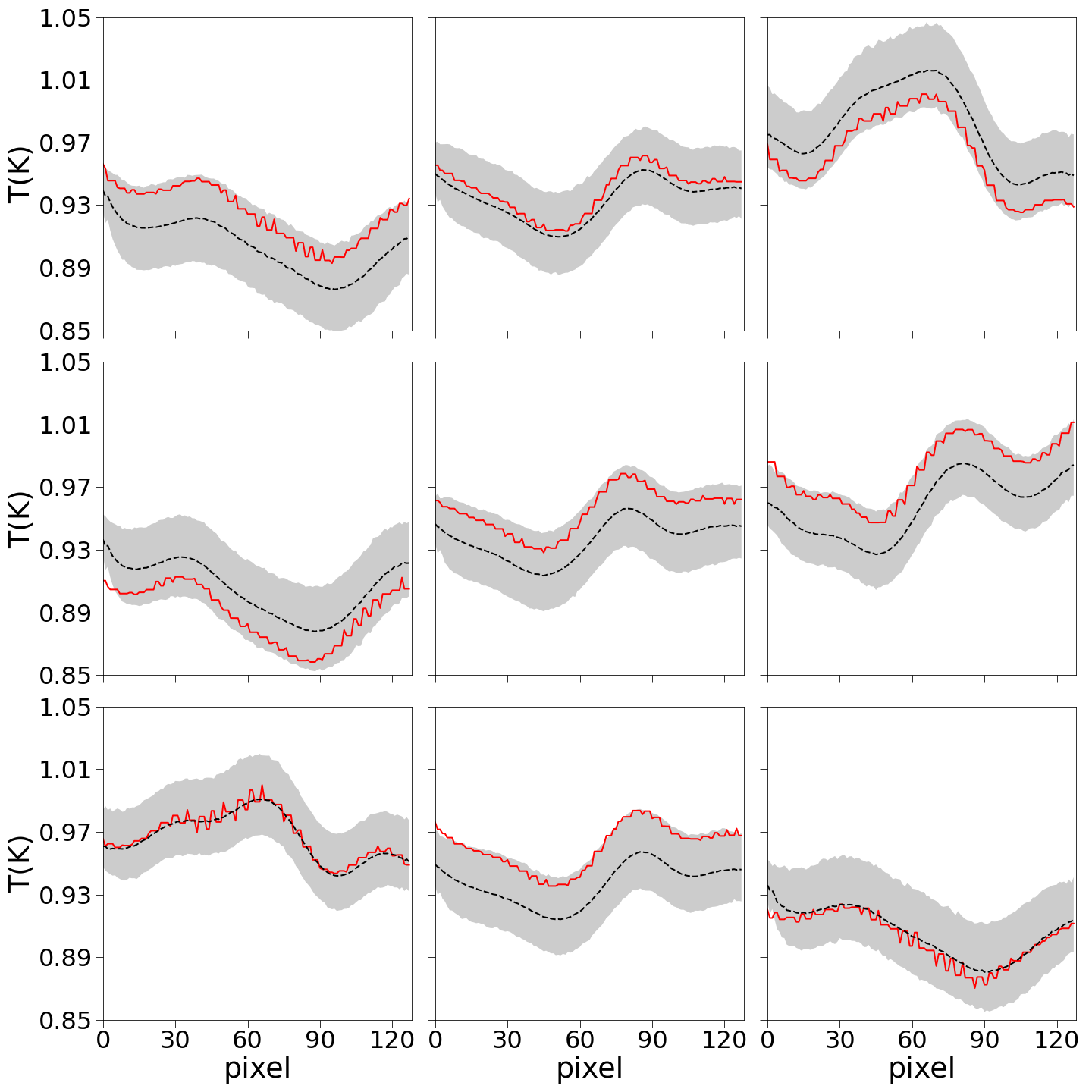}
    \caption{Comparison of the reconstructed maps for randomly selected rows and the true ones. Mean values (black dotted) and 95\% confidence intervals (gray shaded) for individual pixels are obtained using a trained cINN based on 200 realizations of the latent variables $z$. The predicted mean values are within the 95\% C.L. of the true values (red solid) for all 128 pixels of each map.}
    \label{fig:distrubution}
\end{figure*}

Fig.~\ref{fig:distrubution} displays the reconstruction results for randomly selected rows of reconstructed maps. The mean values (black dotted) and 95\% confidence intervals (gray shaded) for individual pixels are obtained by applying the trained cINN, which transforms $p(z)$ to the posterior distribution $p(x|y)$ through the backward mapping process and involves sampling 200 realizations of the latent variables $z$ from the standard normal distribution. As seen, the results show that the predicted mean values are all within the 95\% confidence level when compared with the true values (red solid) across all 128 pixels. Furthermore, the $1\sigma$ of these predictions is approximately 0.01 K, indicating a low variance in the reconstructed maps. The deviations between the reconstructed maps and the true values are also typically of 0.02 K or less, which suggests a good agreement between the inputs and the reconstructed maps.

\subsection{Performance against noise level} \label{S:np}

\begin{figure*}

    \includegraphics[width=\columnwidth]{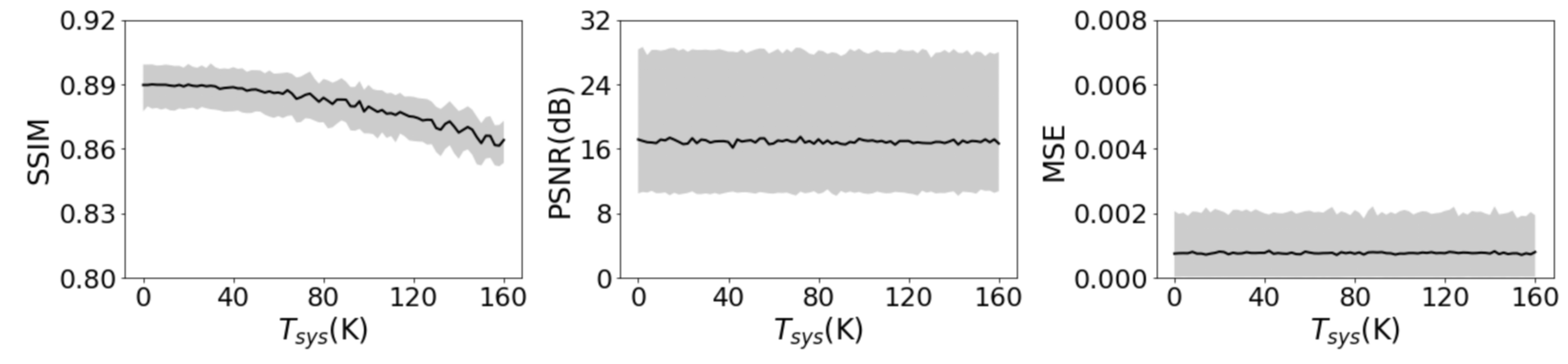}
    \caption{Performance of the cINN reconstruction against white noise, using the SSIM, PSNR, and MSE metrics (from left to right). The calculations are performed on a randomly selected gridding map of TODs, which are centered at the RA and Dec coordinates of $160^\circ$ and $26^\circ$, respectively, at a frequency of 1100 MHz. The mean values (black-solid) and the 95\% C.L. (gray shaded) are estimated by backward mapping of cINN from 200 $z$ realizations.}
    \label{fig:testmse}
\end{figure*}

To thoroughly evaluate the performance of cINN, an investigation is further conducted to determine its ability to produce high-quality reconstructions in the presence of increasing levels of noise present in the input TODs. To do so, in the new test set, we simulate new TODs with a maximum $T_{\rm sys}$ that has been increased from 0 to 160 K. It is important to note that, we did not expand the training samples, instead relying solely on the pre-existing network that was trained with a maximum temperature of $T_{\rm sys}=25$ K. The quality of the reconstructions is evaluated using the PSNR, MSE, and SSIM metrics, as illustrated in Fig.~\ref{fig:testmse}.

As the noise level increases, the values of SSIM show a slight decrease from 0.89 to 0.85 for $T_{\rm sys}$ ranging from 0--160 K, whereas the MSE and PSNR remain roughly unchanged with increasing noise levels. These observations suggest that the performance of cINN is not significantly affected by white noise levels and it has learned the statistical characteristics of the noise during training, demonstrating a strong generalization capability. Our findings support that our cINN model is robust against noise.

\subsection{Time consumption}
We conducted our model training on an NVIDIA Tesla P40 GPU. We have found that the training time of our cINN model is primarily dependent on the number of iterations rather than the number of epochs. When training our network for $10^{4}$ iterations with a batch size of 32, the training process took approximately 1.2 hours, and the GPU memory usage was 2595 MB. For each training run, we executed about $10^4$ iterations until the validation loss increased sharply.

\section{Conclusion} \label{S:con}
In radio observations, the map-making problem--how to reconstruct a plausible sky map from TODs and estimate the signal uncertainty on each pixel--has always been intractable and non-trivial. Unlike the traditional approaches, we  propose to tackle such problem by means of the conditional invertible neural network (cINN). One of the main advantages of our method is that it avoids solving for the ill-posed inverse problem. Moreover, once the network is trained, the reconstruction of the sky map can be performed very fast. 

The use of forward modeling allows for the effortless mapping of the true sky map to TODs, which can incorporate all observational effects, systematics, and data processing. These simulated true sky maps and their associated TODs are then used as a training set and fed into a neural network to train a cINN.  Our cINN model transforms true maps into a latent space and learns the inverse mapping, both of which are conditioned on observations. This joint modeling of the distribution of all pixels provides a comprehensive understanding of the relationship between the true maps and the observations. The trained cINN can then not only reconstruct the true sky map based on a given TOD, but also provide an pixel-by-pixel estimate of the uncertainty.

In order to show the performance of the network, we have performed a simulation of drift-scan observations based on the FAST configuration, which includes 19 beams and covers a frequency range of 1100--1120 MHz. The goal of this study is to initially validate our approach, so for simplicity, we only include white noise and a Gaussian beam response in the simulated TODs, while ignoring other non-ideal effects such as $1/f$ noise and RFIs.

Our method is validated by the test results, which demonstrate high reconstruction accuracy and good agreement between the reconstructed sky maps and the true maps. The test dataset achieves a MSE of $(2.29\pm 2.14) \times 10^{-4}~\rm K^2$, a SSIM of $0.968\pm0.002$, and a PSNR of $26.13\pm5.22$ at the $1\sigma$ level. Furthermore, we observe a slight decrease in the SSIM values as the noise level for $T_{\rm sys}$ increases from 0 to 160 K, ranging from 0.89 to 0.85. However, the MSE and PSNR values remain relatively stable with increasing noise levels.

We have evaluated how underfitting and overfitting affect map reconstruction by comparing checkpoint results above and below our chosen optimal point. Our findings indicate that both cases result in higher MSE values compared with the current point, with underfitting leading to a large uncertainty. Therefore, our current result is optimal. In addition, SSIM and PSNR values do not show any significant deviations from the optimal one, and then MSE appears to be the most sensitive metric in the map reconstruction.

As future work, we aim to validate the cINN approach by incorporating non-ideal observational effects that more accurately reflect real-world scenarios. Furthermore, this framework has the potential to be applied to radio interferometric observations, where imaging can be particularly challenging due to sparse $uv$ coverage.

\normalem
\begin{acknowledgements}
This work is supported by the National Key R\&D Program of China (2018YFA0404502, 2018YFA0404504, 2018YFA0404601, 2020YFC2201600), the Ministry of Science and Technology of China (2020SKA0110402, 2020SKA0110401, 2020SKA0110100), National Science Foundation of China (11890691, 11621303, 11653003, 12205388, 11633004, 11821303), the China Manned Space Project with No. CMS-CSST-2021 (A02, A03, B01), the Major Key Project of PCL, the 111 project No. B20019, and the CAS Interdisciplinary Innovation Team (JCTD-2019-05), the MOST inter-government cooperation program China-South Africa Cooperation Flagship project (grant No. 2018YFE0120800), the Chinese Academy of Sciences (CAS) Frontier Science Key Project (grant No. QYZDJ-SSW-SLH017), and the CAS Strategic Priority Research Program (grant No.XDA15020200).
\end{acknowledgements}
  
\bibliographystyle{raa}
\bibliography{main}

\end{document}